\documentclass[aps,pra,twocolumn]{revtex4}
\newcommand {\be}{\begin{equation}}
\newcommand {\ee}{\end{equation}}
\newcommand {\bey}{\begin{eqnarray}}
\newcommand {\eey}{\end{eqnarray}}
\newcommand {\mydoi}[2]{\href{http://dx.doi.org/#2}{#1}}
\usepackage{epsfig}
\usepackage{hyperref}

\begin{document}

\title{Dynamics of a qubit as a classical stochastic process with time-correlated noise:
minimal measurement invasiveness}

\author{Alberto Montina}
\affiliation{Perimeter Institute for Theoretical Physics, 31 Caroline Street North, Waterloo, 
Ontario N2L 2Y5, Canada}

\date{\today}

\begin{abstract}
So far it has been shown that the quantum dynamics cannot be described as a 
classical Markov process unless the number of classical states is uncountably 
infinite. In this Letter, we present a stochastic model with time-correlated 
noise that exactly reproduces any unitary evolution of a qubit and requires 
just four classical states. The invasive updating of only one bit during a 
measurement accounts for the quantum violation of the Leggett-Garg 
inequalities. Unlike in a pilot-wave theory, the stochastic forces governing 
the jumps among the four states do not depend on the quantum state, but only 
on the unitary evolution. This model is used to derive a local hidden variable 
model, augmented by one bit of classical communication, for simulating entangled 
Bell states.
\end{abstract}
\maketitle

It is a well-established fact that the quantum dynamics among a finite set of
mutually exclusive alternatives (like the {\it up} and {\it down} states of
a $1/2$-spin) cannot be reduced to a Markov process among a 
finite set of classical states. Indeed a time-homogeneous Markov process on a 
finite space always relaxes into a stationary probability distribution, as a 
consequence of the Perron-Frobenius theorem~\cite{perron}. The impossibility 
of exactly simulating a quantum system through a Markov process on a finite 
classical space is also a consequence of a theorem proved in Ref.~\cite{montina}.
There we showed
that the dimension of the classical space cannot be smaller than the quantum 
state manifold dimension under the hypothesis of Markov dynamics, that is,
the classical space must be uncountably infinite. Pilot wave theories provide
an example of this overflow of classical resources. Indeed in their framework
the dynamics among the set of alternatives explicitly depends on 
the quantum state, which actually turns to be part of the classical description. 
Pilot wave theories of finite-dimensional quantum 
systems were considered for example in Refs.~\cite{aaronson,hardy}. While
they reintroduce a classical realistic picture of the quantum world, they
are unavoidably characterized by a feature that is absent in prequantum 
physics, namely, the invasiveness of measurements. Thus, measurements
do not provide a mere updating of knowledge about the actual state, but
intrinsically introduce a perturbation on the system. In every known classical 
model of quantum dynamics this perturbation demands an invasive updating of an 
uncountably infinite amount of information. For example, in the case of a qubit,
two continuous real variables need to be updated.

The Leggett-Garg inequalities provide a useful test for deciding if a set of data 
can be explained by a measurement-noninvasive classical theory~\cite{legg}. Indeed 
they are violated by quantum mechanics. These inequalities are analogous to the Bell 
inequalities for the Einstein-Podolsky-Rosen (EPR) experiment~\cite{bell} with the 
local measurements being replaced by two consecutive measurements on a two-state 
quantum oscillator. This analogy suggests an interesting question. On the one hand 
the violation of the Leggett-Garg inequalities demands the invasiveness of measurements 
in any classical theory of two-state quantum oscillators. On the other hand, the 
violation of the Bell inequality implies that a 
classical simulation of Bell correlations requires some communication between the 
parties. It is known that a finite amount of classical communication, namely one 
bit, can actually account for this violation~\cite{bacon}. Thus it is natural to 
wonder if it is possible to simulate the quantum dynamics by a classical model that 
needs the invasive updating of a finite amount of information. 

Generalizing the case discussed in Ref.~\cite{legg}, in this Letter we consider the 
scenario where two consecutive measurements of the same observable are performed, at 
times $t_0$ and $t_1>t_0$, on a qubit undergoing a generic unitary evolution. The 
outcome of each measurement is one of two orthogonal states, denoted by $|\pm1\rangle$. 
They can be, for example, the left and right states of a double-well system~\cite{legg}. 
We show that this scenario can be classically simulated by a two-bit classical model 
with time-correlated noise. The key ingredients of our model are {\it time correlation} 
of the noise and {\it minimal measurement invasiveness}. The first ingredient is 
required for circumventing the constraint on the classical space dimension proved in 
Ref.~\cite{montina}, as this constraint does not hold for non-Markov processes. The 
second ingredient accounts for the invasiveness of measurements by demanding that only 
one bit is invasively updated. Unlike in a pilot-wave theory, the noise governing the 
dynamics is independent of the measurement times $t_0$ and $t_1$ and depends only on the 
unitary evolution. In particular, the noise is independent of the quantum state. This 
feature and the finite number of classical states distinguish our result from the previous 
ones, such as the class of hidden variable theories considered in Ref.~\cite{aaronson}, 
where the stochastic matrices are supposed to depend on the quantum state.
At first glance, the finiteness of the classical space seems paradoxical 
since the quantum state space is infinite. Indeed, the quantum information is not 
encoded into the classical state of a single execution, but into 
the statistical behavior of many executions. 

Like in a pilot-wave theory,
in the presented model the system is supposed to be, at any instant, definitely in one of the 
two orthogonal states $|\pm1\rangle$. More precisely, the model contains a two-value
discrete index, $s=\pm1$, that determines the outcome of a measurement on the basis
$\{|-1\rangle,|1\rangle\}$. If at some time the index $s$ is equal to some value 
$s_0$ and a measurement is performed, then the outcome is $|s_0\rangle$. Furthermore, 
the measurement does not change the subsequent value of the index $s$. The role of $s$ 
is similar to the role played by the position variables in the de Broglie-Bohm mechanics. 
To derive the stochastic model, we first introduce a simple measurement-noninvasive model 
that captures some features of a qubit. The qubit is described by just the 
bit $s$ that is kicked by a time-correlated noise depending on the unitary evolution 
(first ingredient). Since this model is measurement-noninvasive, it satisfies
the Leggett-Garg inequalities and it is not equivalent to quantum mechanics.
We then modify the model by introducing another bit that is invasively updated by 
the measurements (second ingredient). The dynamics of $s$ is ruled by both the 
noise and the additional bit. It is shown that this minimal improvement is sufficient 
for exactly reproducing the quantum transition between two consecutive measurements.

It is useful to represent the quantum states by Bloch vectors. The unitary
evolution is described by a rotation on the Bloch sphere. We denote by 
$R(t_a,t_b)$ the rotation operator along any time interval $[t_a,t_b]$
and by vectors $\pm\vec n$ the states $|\pm1\rangle$. The quantum probability of 
having $|s_1\rangle$ at time $t_1$, given $|s_0\rangle$ at time $t_0$, is
\bey\label{quantum_prob}
P_Q(s_1,t_1|s_0,t_0)&=&\frac{1}{2}[1+s_1 s_0 \vec n\cdot\vec v], \\
\label{def_v}
\vec v&\equiv& R(t_0,t_1)\vec n.
\eey

Let us introduce the measurement-noninvasive model. First, we define the noise 
and the rule governing the jumps of $s$. Then, we derive the transition 
probability of $s$ along a time interval. The noise variable is a unit 
vector $\vec x(t)$ that is a function of time. We denote by $\rho[\vec x(t)]$ 
the marginal probability distribution of $\vec x(t)$ at time $t$ and by 
$\rho(\vec x)$ the probability distribution of the function $\vec x$ 
[i.e., $\rho(\vec x)$ is a functional].
The statistical distribution of the noise, namely $\rho(\vec x)$, is defined by 
the equations
\bey\label{margin_noise}
\rho[\vec x(t_a)]=(4\pi)^{-1}, \\
\label{evo_noise}
\vec x(t_b)=R(t_a,t_b)\vec x(t_a)
\eey
for any $t_a$ and $t_b$. The first equation gives the 
marginal probability distribution of $\vec x(t_a)$ at time $t_a$. The second 
equation establishes a deterministic relation between the value of the noise 
variable at different times. The procedure for generating each realization of
the noise is as follows. First, we generate a random vector $\vec x(t_a)$ at
some time $t_a$ according to Eq.~(\ref{margin_noise}). Then, we determine
$\vec x(t)$ at any time by using Eq.~(\ref{evo_noise}). The noise
is clearly time-correlated, that is, the correlation function
$\langle x_i(t_a) x_j(t_b)\rangle$ is not equal to zero for $t_a\ne t_b$.
Given the noise function $\vec x(t)$, we need a rule for the dynamics of
$s(t)$. We employ the simplest deterministic rule by assuming that the index $s(t)$ 
undergoes a jump whenever $\vec n\cdot\vec x(t)$ changes the sign,
that is, whenever $\vec x(t)$ crosses the geodesic of the Bloch sphere
lying on a plane orthogonal to $\vec n$. 
Let us summarize the noninvasive model. \newline
{\bf Model 1.} {\it Let the unit vector $\vec x(t)$ be a stochastic function
of time, whose statistical property is given by Eqs.~(\ref{margin_noise},
\ref{evo_noise}). In each Monte Carlo execution, the two-value index $s(t)$
undergoes a jump whenever $\vec x(t)\cdot \vec n$ changes the sign. }

Thus, the Monte Carlo procedure for generating the value of $s(t_1)=s_1$ 
at time $t_1$ given $s(t_0)=s_0$ at time $t_0$ is as follows. A noise
function $\vec x(t)$ is generated according to Eqs.~(\ref{margin_noise},
\ref{evo_noise}). If the signs of $\vec x(t_0)\cdot\vec n$ and 
$\vec x(t_1)\cdot\vec n$ are equal (even number of jumps), then
$s_1$ is set equal to $s_0$, otherwise $s_1$ is set equal to $-s_0$. 
The procedure is repeated for each realization.

Notice that all we need to know about the noise is the real function 
$\xi: t\rightarrow \vec n\cdot\vec x(t)$. 
In fact, we could just suppose
that $\xi$ is the noise and regard $\vec x$ as an intermediary 
tool for mapping each unitary evolution to a
statistical distribution of $\xi$, namely, for generating the map
$R\rightarrow\rho(\xi)$ from the function $R:t_a,t_b\rightarrow R(t_a,t_b)$
to the probability distribution (which is a functional) of the noise $\xi$.
As with $\vec v$ and $\vec v(t)$, if not differently indicated, we denote by 
$\xi$ the noise function and by $\xi(t)$ its value at time $t$.
Thus, $F(\xi)$ is meant as a functional of the noise 
at every time and $F[\xi(t)]$ as a function of the noise at time $t$.
Notice that the process $\xi$ is not Markovian,
that is, the marginal probability $\rho(\xi_c|\xi_b,\xi_a)$ of 
having $\xi_c=\xi(t_c)$ at time $t_c$ given $\xi_b=\xi(t_b)$ and 
$\xi_a=\xi(t_a)$ at previous times $t_b$ and $t_a$ is not equal, 
in general, to the marginal probability $\rho(\xi_c|\xi_b)$.

Let us denote by $P_\xi(s_1;t_1|s_0;t_0)$ the transition probability
from $s_0$ at time $t_0$ to $s_1$ at time $t_1$, given a noise
realization $\xi$. The rules defining model~1 imply that
$
P_\xi(s_1;t_1|s_0;t_0)\equiv\theta[s_1 s_0\vec n\cdot\vec x(t_1)\vec n\cdot\vec x(t_0)],
$
$\theta$ being the Heaviside function.
Using Eqs.~(\ref{def_v}) and (\ref{evo_noise}), we have that
\be\label{cond0}
\begin{array}{c}
P_\xi(s_1;t_1|s_0;t_0)=
\theta[s_1 s_0\vec n\cdot\vec x(t_1)
\vec v\cdot\vec x(t_1)],
\end{array}
\ee
From Eq.~(\ref{cond0}) and the statistical distribution of $\vec x(t_1)$
defined by Eq.~(\ref{margin_noise}), we find, by
the marginalization over $\vec x(t_1)$, that the probability of
the transition $s_0\rightarrow s_1$ is
\be\begin{array}{l}
P(s_1,t_1|s_0,t_0)=\frac{1}{4\pi}\int d^2x \theta[s_1 s_0 
(\vec n\cdot\vec x)(\vec v\cdot\vec x)]= 
\vspace{2mm}   \\
1-\frac{1}{\pi}\arccos(s_1 s_0\vec n\cdot\vec v),
\end{array}
\ee
Thus, the model does not exactly reproduce the quantum probability given 
by Eq.~(\ref{quantum_prob}). For example, in the case of Rabi oscillation 
between states $|\pm1\rangle$, the quantum probability is a cosine squared 
function of $\omega(t_1-t_0)$, $\omega$ being the Rabi frequency. Conversely, 
the model presented here gives a triangle function. In particular, for a 
small evolution time the classical probability scales as $t_1-t_0$, 
whereas the quantum probability scales as $(t_1-t_0)^2$. While the model 
is not exact, it has the nice property of generating an oscillatory dynamics, 
which a Markov process on a finite set of states fails to give. This property 
is granted by the time correlation of the noise.

Just as a local model satisfies the Bell inequalities, this
measurement-noninvasive model satisfies the Leggett-Garg inequalities, 
which are violated by quantum systems. To simulate exactly the quantum 
transition between two measurements we need to introduce some additional 
variable that is invasively updated by the first measurement. 
We now present an exact model that uses just one additional bit, denoted 
by a discrete index, $r(t)$, taking the values $\pm1$. 
The model of a qubit is as follows. \newline
{\bf Model 2.} {\it Let the unit vectors $\vec x_1(t)$ and $\vec x_{-1}(t)$ 
be two stochastic functions of time. They are statistically independent and 
the statistical property of each function is given by 
Eqs.~(\ref{margin_noise},\ref{evo_noise}). 
The qubit is described by two indices $s(t)=\pm1$ and $r(t)=\pm1$, which are 
functions of time $t$. The index $s(t)$ is directly measurable at any time and 
is not modified by a measurement, whereas $r(t)$ is invasively updated. If
a measurement of $s(t_0)\equiv s_0$ is performed at time $t_0$, then the 
index $r(t_0)$ is set equal to 
\be\label{rule_r}
r_0=\text{\rm sign}\left\{[\vec x_1(t_0)\cdot\vec n]^2-
			[\vec x_{-1}(t_0)\cdot\vec n]^2\right\},
\ee 
The index $r(t)$ remains constant after the measurement, whereas $s(t)$ undergoes a 
jump whenever $\vec x_{r_0}(t)\cdot\vec n$ changes the sign. A second measurement
reveals the value of $s(t_1)\equiv s_1$ at time $t_1>t_0$. }

\begin{figure}
\epsfig{figure=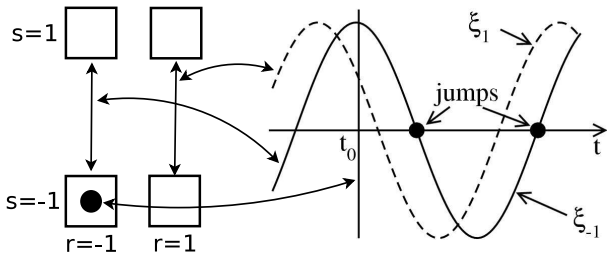,width=8.cm}
\caption{Schematic representation of the $4$-state model. During
a unitary evolution the index $s$ undergoes jumps whenever $\xi_r$ 
changes sign, whereas $r$ remains constant. For example, if a measurement
is made at time $t_0$ and $r$ is set equal to $-1$, the jumps occur
at the filled circles.}
\label{fig1}
\end{figure}

Like in the previous model, all we need to know about the noise are the functions 
$\xi_{\pm1}: t\rightarrow \vec n\cdot\vec x_{\pm1}(t)$.
In the following we will denote by $\xi$ the pair of noise functions $\xi_{\pm1}$. 
A schematic representation of the model for a particular realization of the noise 
is given in Fig.~\ref{fig1}. Notice in figure that a measurement at time $t_0$ sets 
$r=-1$, since $\xi_{-1}^2(t_0)>\xi_1^2(t_0)$, in accordance with Eq.~(\ref{rule_r}).

For each noise realization $\xi$, 
the probability of having outcome $s_1$ at time
$t_1$, given outcome $s_0$ at time $t_0$, is equal to 
$
P_{\xi}(s_1;t_1|s_0;t_0)=
\theta[s_1 s_0\hspace{1mm}\vec n\cdot\vec x_{r_0}(t_1)\hspace{1mm}
\vec n\cdot\vec x_{r_0}(t_0)]
$
where $r_0$ is given by Eq.~(\ref{rule_r}).
Thus, using Eq.~(\ref{def_v}) and (\ref{evo_noise}), we have that
\be
\label{dynam_rule}
P_\xi(s_1;t_1|s_0;t_0)\equiv
\theta[s_1 s_0\hspace{1mm}\vec n\cdot\vec x_{r_0}(t_1)\hspace{1mm}
\vec v\cdot\vec x_{r_0}(t_1)].
\ee
Similarly, Eq.~(\ref{rule_r}) can be written, 
through Eq.~(\ref{def_v}) and (\ref{evo_noise}), as
\be\label{rule_r2}
r_0=\text{\rm sign}\left\{[(\vec x_1(t_1)\cdot\vec v]^2-
			[\vec x_{-1}(t_1)\cdot\vec v]^2\right\},
\ee 

Like in the previous model, the probability $P(s_1;t_1|s_0;t_0)$ for the 
transition from $s_0$ at time $t_0$ to $s_1$ at time $t_1$ is obtained by 
averaging over the noise realizations. Thus, from 
Eqs.~(\ref{margin_noise},\ref{dynam_rule},\ref{rule_r2}) we have that
\be\begin{array}{c}
P(s_1;t_1|s_0;t_0)=\frac{1}{(4\pi)^2}\sum_{r=\pm1}\int d^2 x_1 d^2 x_{-1}
\vspace{1.5mm}\\
\theta(s_1 s_0\hspace{0.5mm}\vec n\cdot\vec x_{r}\vec v\cdot\vec x_{r}) 
\theta\{r [(\vec x_1\cdot\vec v)^2-
			(\vec x_{-1}\cdot\vec v)^2]\}.
\end{array}
\ee
Noting that the two terms in the sum over $r$ give the same contribution,
it is not difficult to show that 
\be
P(s_1;t_1|s_0;t_0)=\int d^2 x
\theta(s_1 s_0\hspace{0.5mm}\vec n\cdot\vec x\hspace{0.5mm}
\vec v\cdot\vec x) I(\vec v\cdot\vec x)
\ee
with $I(\eta)\equiv\frac{1}{8\pi^2} \int d^2 y
\theta[\eta^2-(\vec v\cdot\vec y)^2]=\frac{1}{2\pi}|\eta|$.
This gives the equation
\be\label{inte_ks}
P(s_1;t_1|s_0;t_0)=\int d^2x\theta(s_0\vec n\cdot\vec x)
\rho_{ks}(\vec x;s_1\vec v),
\ee
where $\rho_{ks}(\vec x;\vec w)\equiv\frac{1}{\pi}
\vec w\cdot\vec x\hspace{0.5mm}\theta
(\vec w\cdot\vec x)$ is the probability distribution associated
with quantum state $\vec w$ in the Kochen-Specker (KS) model~\cite{ks}.
The integral in Eq.~(\ref{inte_ks}) is well-known \cite{ks} and 
gives the quantum probability of having the state $s_0\vec n$,
given the state $s_1\vec v$ and viceversa. Thus, we have proved 
that 
$$P(s_1;t_1|s_0;t_0)=P_Q(s_1;t_1|s_0;t_0),$$ that is, the
stochastic model exactly reproduces the quantum transition between
two measurements.
Unlike in a pilot-wave theory, the noise $\xi$
does not depend on measurement times $t_1$ and $t_2$. Thus, the noise 
value $\xi(t)$ at any time $t$ is independent of the quantum state at 
that time. Indeed the information on the quantum state is 
encoded in the correlation between $(s,r)$ and $\xi$.

There is a close relation between this model and some results in quantum 
communication. Let us consider the following EPR scenario~\cite{bell}. 
Two spatially separate parties, Alice and Bob, each receive one of two 
maximally entangled qubits. Alice performs a local projective measurement 
on the single-qubit basis $(\vec v_1,-\vec v_1)$, while Bob on the basis 
$(\vec v_0,-\vec v_0)$. The two-value indices $s_0$ and $s_1$ are 
defined so that Alice and Bob's outcoming states are the Bloch vectors 
$-s_1\vec v_1$ and $s_0\vec v_0$, respectively.
With a suitable choice of the reference frame on the Bloch spheres, 
the joint probability distribution of $s_0$ and $s_1$ is
$
P_e(s_0,s_1|\vec v_0,\vec v_1)=\frac{1}{4}[1+s_0 s_1 \vec v_0\cdot\vec v_1].
$
According to Bell's theorem, a local hidden variable model 
cannot reproduce this probability distribution and some 
post-measurement communication between the
parties has to be exchanged. How much communication is required?
In Ref.~\cite{brassard} it was shown that a finite amount 
of communication, namely $8$ bits, is sufficient for reproducing 
the Bell correlations.
This result was improved in Ref.~\cite{bacon}, where it was shown
that an exact simulation demands a communication of just one bit. 
An alternative model with minimal communication was derived in 
Ref.~\cite{montina2} from the Kochen-Specker model~\cite{ks}. 
The common setting of a one-way classical 
protocol for simulating entanglement is as follows. 
Bob and Alice share a random variable $X$. Given the 
measurement $(\vec v_0,-\vec v_0)$, Bob generates the outcome 
$s_0$ and an additional discrete index $m$ according
to a probability distribution that depends on $\vec v_0$
and the shared variable $X$. Then he sends $m$ to Alice.
Alice generates the outcome $s_1$ of the measurement 
$(\vec v_1,-\vec v_1)$ with a probability that depends on both
$\vec v_1$, $X$ and the communicated index $m$. A stochastic model 
of quantum dynamics on a finite classical space, such as that introduced 
in this paper, can be easily converted into a model of entanglement, 
where the stochastic noise and the classical state play the role of 
$X$ and the communicated information, respectively.

Let us show that the stochastic model of qubit derived here
can be converted into the model of entanglement reported 
in Ref.~\cite{montina2}. Suppose that at time $t_0$ the
qubit is in the mixture 
$\frac{1}{2}(|1\rangle\langle1|+|-1\rangle\langle-1|)$
and a projective measurement is performed on the basis 
$(\vec n,-\vec n)$. The probability distribution of 
outcome $s_0$ is $\rho(s_0)=\frac{1}{2}$. Then the qubit undergoes 
two consecutive unitary evolutions along the time intervals $[t_0,t]$ 
and $[t,t_1]$. At time $t_1$ another measurement on the basis 
$(\vec n,-\vec n)$ is made and it gives outcome $s_1$.
Each measurement at time $t_i$, with 
$i=1,2$, is actually equivalent to a measurement on the basis 
$(\vec v_i,-\vec v_i)$ at the same time $t$, where 
$\vec v_i\equiv R(t_i,t)\vec n$. Thus, we have that the joint 
probability of $s_0$ and $s_1$ is formally equal to the probability 
distribution $P_e(s_0,s_1|\vec v_0,\vec v_1)$ for two entangled
qubits in the EPR scenario.
Indeed, both the marginal distributions and the correlations of
$s_0$ and $s_1$ are identically reproduced.
From our stochastic model we find that, given the vectors
$\vec x_i(t)\equiv\vec y_i$, the joint probability distribution of 
$(s,r)$ at time $t$ and outcome $s_0$ at time $t_0$ is
\be\begin{array}{c}
\label{prep}
P_B(s_0,s,r|y;\vec v)= 
\vspace{1mm}\\
\frac{1}{2}\theta[(s_0\vec v\cdot\vec y_r)(s \vec n\cdot\vec y_r)]
\theta\{r[(\vec v\cdot\vec y_1)^2-(\vec v\cdot\vec y_{-1})^2]\},
\end{array}
\ee
where 
$y\equiv(\vec y_1,\vec y_{-1})$. Similarly, we have that, given $y$ and 
$(s,r)$, the probability of the outcome $s_1$ at time $t_1$ is
\be
\label{meas}
P_A(s_1|s,r,y;\vec v_1)=
\theta[(s\hspace{1mm} \vec n\cdot \vec y_r)(s_1\hspace{1mm}\vec v_1\cdot\vec y_r)].
\ee
Finally, the joint probability of $s_0$ and $s_1$ is 
\be\label{equiv}
\begin{array}{c}
P_e(s_0,s_1|\vec v_0,\vec v_1)=  
\vspace{1mm}\\
\sum_{s,r}\int d^4y P_A(s_1|s,r,y;\vec v_1)
P_B(s_0,s,r|y,\vec v_0)\rho(y).
\end{array}
\ee
These three equations give a model of entanglement where $y$ is the 
shared noise and $(s,r)$ the communicated bits. Notice that $\vec n$ 
is just a free parameter of the model. It can be eliminated by the 
transformation 
$s\rightarrow s\hspace{1mm}\text{sign}(\vec n\cdot\vec y_r)$. 
Furthermore, the marginal probability distribution of $s$ after the 
transformation is uniform and independent from $\vec v$. Thus, $s$ can 
be included in the set of shared variables and, indeed, incorporated 
in $y$. In this way we obtain a local hidden variable model of entanglement,
augmented by one bit of communication (namely, $r$), given by the conditional
probabilities
$$
\begin{array}{c}
P_B(s_0,r|y;\vec v_0)= 
\theta(s_0\vec v_0\cdot\vec y_r)
\theta\{r[(\vec v_0\cdot\vec y_1)^2-(\vec v_0\cdot\vec y_{-1})^2]\},
\vspace{2mm} \\
P_A(s_1|r,y;\vec v_1)=
\theta(s_1 \hspace{1mm}\vec v_1\cdot\vec y_r).
\end{array}
$$
This is the model derived in Ref.~\cite{montina2}, set in a slightly
different form. The process
can be reverted and one can obtain the stochastic model directly
from the model of entanglement. 

In conclusion, we have presented a stochastic model with 
time-correlated noise that exactly reproduces any unitary
evolution of a qubit by using just $4$ classical states.
The time correlation of the noise allowed us to overcome
the constraint of the theorem proved in Ref.~\cite{montina} on
the dimensionality of the classical space. A generalization to 
higher-dimensional quantum systems can have some interesting 
implications. First, it would automatically give a local
hidden variable model of entanglement, augmented by a finite 
amount of one-way communication, for $n$ Bell states.
Apart from some approximate protocols reported in 
Ref.~\cite{montina2}, such a model is at present missing. 
Second, this generalization can suggest more efficient methods for 
simulating the dynamics of high-dimensional quantum systems.
At first glance, this approach does not seem to provide any 
computational benefit. Indeed, the evaluation of each noise 
realization requires one to solve the Schr\"odinger equation, 
thus it is not less complicated than directly solving the dynamics
of the quantum state. However, we have seen that only
partial information about the noise (in our model,
the real functions $\xi_i$) is actually involved in the 
dynamics of the discrete indices. Thus, one could envisage a
computational strategy for computing, exactly or with some 
approximation, this partial information without passing through 
the Schr\"odinger equation. \newline
{\it Acknowledges.} I thank Cecilia Flori for stylistic suggestions.
I am grateful for useful discussions with Caslav Brukner, Sandru
Popescu, Paul Busch, Terry Rudolph and Jonathan Barrett.
Research at Perimeter Institute for Theoretical Physics is
supported in part by the Government of Canada through NSERC
and by the Province of Ontario through MRI.


\begin{thebibliography}{20}
\bibitem{perron} R. B. Bapat and T. E. S. Raghavan, 
``Nonnegative matrices and applications'', (Cambridge 
University Press, 1997); J. Ding and A. Zhou, ``Nonnegative matrices, positive
operators, and application'' (Hackensack, N.J., World Scientific, 2009).
\bibitem{montina}
\mydoi{A. Montina, Phys. Rev. A {\bf 77}, 022104 (2008)}
{10.1103/PhysRevA.77.022104};
\mydoi{A. Montina, Phys. Rev. A {\bf 83}, 032107 (2011)}
{10.1103/PhysRevA.83.032107}.
\bibitem{aaronson} 
\mydoi{S. Aaronson, Phys. Rev. A {\bf 71}, 032325 (2005)}
{10.1103/PhysRevA.71.032325}.
\bibitem{hardy} 
\mydoi{L. Hardy et al., Phys. Rev. A {\bf 45}, 4267 (1992)}
{10.1103/PhysRevA.45.4267}.
\bibitem{legg} 
\mydoi{A. J. Leggett and A. Garg, Phys. Rev. Lett. {\bf 54}, 857 (1985)}
{10.1103/PhysRevLett.54.857}.
\bibitem{bell} J. S. Bell, ``Speakable and unspeakable in quantum mechanics'',
(Cambridge University Press, 2004).
\bibitem {bacon} 
\mydoi{B. F. Toner and D. Bacon, Phys. Rev. Lett. {\bf 91}, 187904 (2003)}
{10.1103/PhysRevLett.91.187904}.
\bibitem{ks} 
S. Kochen and E. Specker, J. Math. Mech. {\bf 17}, 59 (1967).
\bibitem{brassard} 
\mydoi{G. Brassard, R. Cleve, A. Tapp, Phys. Rev. Lett. {\bf 83}, 1874 (1999)}
{10.1103/PhysRevLett.83.1874}.
\bibitem{montina2} 
\mydoi{A. Montina, Phys. Rev. A {\bf 84}, 042307 (2011)}
{10.1103/PhysRevA.84.042307}.

\end{thebibliography}
\end{document}